\documentclass[aps,pra,twocolumn,superscriptaddress]{revtex4-1}

\usepackage{amsmath}
\usepackage{graphicx}
\usepackage{amsfonts}
\usepackage{color}
\usepackage{upgreek}
\usepackage{mathtools}
\usepackage{hyperref}
\usepackage{comment}
\hypersetup{
    colorlinks = true,
    linkcolor =blue,
	citecolor=blue, 
	urlcolor=blue 
}

\begin{document}

\title{Interaction-induced  transition in quantum many-body detection probability}

\author{Archak Purkayastha}
\email{archak.p@phy.iith.ac.in}
\affiliation{Department of Physics, Indian Institute of Technology, Hyderabad, 502284, India}
\affiliation{Centre for complex quantum systems, Aarhus University, Nordre Ringgade 1, 8000 Aarhus C, Denmark}

\author{Alberto Imparato}
\email{imparato@phys.au.dk}
\affiliation{Department of Physics and Astronomy, Aarhus University, \\ Ny Munkegade, Building 1520, DK–8000 Aarhus C, Denmark}

\date{\today}

\begin{abstract}
With the advent of digital and analog quantum simulation experiments, it is now possible to experimentally simulate dynamics of quantum many-body lattice systems and make site-resolved measurements. These experiments make it pertinent to consider the probability of getting any specific measurement outcome, which we call the `signal', on placing multiple detectors at various sites while simulating dynamics of a quantum many-body lattice system. In this work, we formulate and investigate this problem, introducing the concept of quantum many-body detection probability (QMBDP), which refers to the probability of detecting a chosen signal at least once in a given time. We show that, on tuning some Hamiltonian parameters, there can be sharp transition from a regime where QMBDP $\approx 1$, to a regime, where QMBDP $\approx 0$. Most notably, the effects of such a transition can be observed at a single trajectory level. This is not a measurement-induced transition, but rather a non-equilibrium transition reflecting opening of a specific type of gap in the many-body spectrum. We demonstrate this in a single-impurity non-integrable model, where changing the many-body interaction strength brings about such a transition. Our findings suggest that instead of measuring expectation values, single-shot stroboscopic measurements could be used to observe non-equilibrium transitions.
\end{abstract}

\maketitle

{\it Introduction---}
A fundamental question relevant across various branches of science is whether a chosen type of signal can be detected at a given position. One of the oldest mathematical formulations of the problem concerns a particle undergoing random walk. The seminal P{\'o}lya's theorem \cite{Polya_1921} states that in one and two dimensions, the particle will eventually be detected with certainty regardless of the position of the detector, while in three dimensions there is a finite chance that the particle is never detected. Similar questions have been extensively studied in complex classical stochastic systems under the guise of survival and first-passage probabilities \cite{Alan_2013, Redner_2022}. In the realm of quantum systems, these questions have been considered from the perspective of time-of-arrival and quantum search problems \cite{Kulkarni_2023,Liu_2023, Liu_2022, Liu_2020, Dubey_2021,  Thiel_2021,Thiel_2020_2,Thiel_2020_1,Yin_2019,Thiel_2018,Friedman_2017,
Chakraborty_2016, Dhar_2015_1, Dhar_2015_2, Grunbaum_2013, Anastopoulos_2012,Krovi_2006,Bach_2004, Damborenea_2002, Muga_2000}. Most of these studies have primarily focused on single-particle systems with a single detector placed at a given location. Over the past decade, digital and analog quantum simulation experiments \cite{Schafer_2020,Morgado_2021,Monroe_2021,Daley_2022,Tran_2023} have been developed to simulate the dynamics of quantum many-body lattice systems and make site-resolved measurements, for example, with quantum gas microscopes \cite{Bakr_2009, Sherson_2010, Gross_2021, Wei_2022, Hilker_2017, Brown_2017, Cheuk_2016, Boll_2016, Maxwells_2016, Greif_2016, Cheuk_2016, Preiss_2015, Omran_2015, Parsons_2015, Fukuhara_2015, Haller_2015, Islam_2015, Cheneau_2012}.  The advent of these experiments makes it pertinent to consider the detection probability of a signal in the presence of quantum many-body interactions, with multiple detectors placed at different lattice sites. In this paper, we formulate and investigate this problem, providing an interesting example.

We define the signal as a particular measurement outcome of simultaneous stroboscopic projective measurements by the detectors. We introduce the notion of quantum many-body detection probability (QMBDP), by which we refer to the probability that the signal is detected at least once within a given time.
Choosing a single-impurity non-integrable model \cite{Santos_2011,Brenes_2018,Brenes_2020_1,Brenes_2020_2,Bertini_2021}, we demonstrate that, depending on initial state, there can be a sharp transition in QMBDP over a finite but large regime of time. In our chosen model, such a transition is brought about by tuning the many-body interaction strength. This transition is from a regime where the signal is almost certainly detected (QMBDP $\approx 1$), to a regime where the signal almost certainly not detected (QMBDP $\approx 0$). 
This is not a class of measurement induced phase transition \cite{Buchhold_2021,Tang_2020,Skinner_2019, Dhar_2016}. Instead, as we show in general, such a transition is related to opening a specific type of gap in the many-body spectrum of the system. It can be explained via an unconventional application of van Vleck perturbation theory (VVPT). It also manifests in far-from-equilibrium dynamical properties in absence of the detectors, for example, in domain-wall dynamics. However, we find that, the transition in QMBDP is much sharper than that in other dynamical properties. 
Most interestingly, since QMBDP takes into account the effects of measurement backaction, a transition in QMBDP can be captured at a single trajectory level. This opens the possibility of observing non-equilibrium transitions via single-shot stroboscopic measurements, rather than via obtaining dynamics of expectation values. This fact is both fundamentally interesting and experimentally appealing, with potential technological implications.

{\it Introducing QMBDP and our example---}
Consider a quantum many-body lattice system with Hilbert space dimension $D$ in a state far-from-equilibrium. Suppose that some detectors are placed at some specific sites, which are switched on in stroboscopic steps of time $\tau$. The detectors make instantaneous projective measurement of some observable, say particle number, in those sites. In this situation, one can ask about the probabilities of making a chosen type of observation. For example, if there are two particle detectors, one can ask, what is the probability that they click simultaneously. We can think of the chosen type of observation as the `signal'. 
 Let the Hamiltonian for the lattice system be $\hat{H}$, the initial state be $\hat{\rho}(0)$, the projection operator corresponding to measurement of the `signal' be $\hat{P}$, and the complementary projection operator be $\hat{Q}=\hat{\mathbb{I}}-\hat{P}$, where $\hat{\mathbb{I}}$ is the identity operator. Using Born rule and a little algebra, the probability of not detecting the signal in $n$ steps is
\begin{equation}
\label{def_Rn_MQ}
R_n(\tau) = {\rm Tr}\left(\left[\hat{M}_Q(\tau)\right]^n \hat{A}(\tau) \hat{\rho}(0) \hat{A}^\dagger(\tau) \left[\hat{M}_Q^{\dagger}(\tau)\right]^n\right),
\end{equation} 
where, $\hat{M}_Q(\tau) = \hat{Q} e^{-i\hat{H}\tau} \hat{Q},~~\hat{A}(\tau)=\hat{Q}e^{-i\hat{H}\tau} \hat{P} + \hat{M}_Q(\tau)$.
We call $R_n(\tau)$ no-detection probability. This is the analog of `survival probability' studied in classical stochastic systems \cite{Alan_2013,Redner_2022}. The QMBDP, i.e, the probability that the signal is detected at least once within time $n\tau$, is given by 
$T_n(\tau) = 1-R_n(\tau)$.

\begin{figure}
\includegraphics[width=\columnwidth]{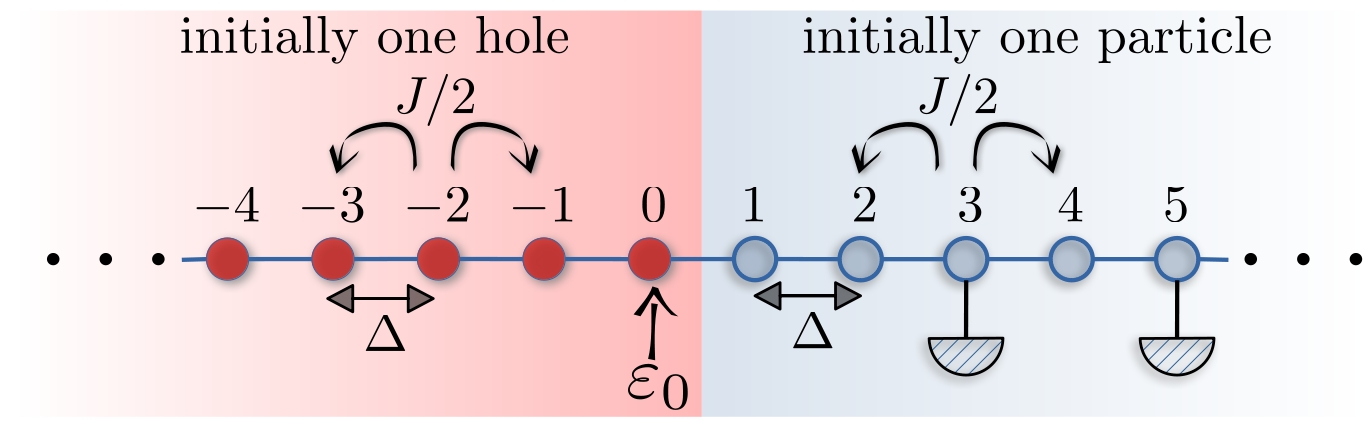}
\caption{We consider the Hamiltonian $\hat{H}=-\sum_{\ell=-N/2+1}^{N/2-1} \left[\frac{J}{2}\left( \hat{c}_\ell^\dagger \hat{c}_{\ell + 1} +  \hat{c}_{\ell + 1}^\dagger \hat{c}_\ell \right) + \Delta \hat{n}_\ell \hat{n}_{\ell + 1} \right]+\varepsilon_0 \hat{n}_{0}$, where $\hat{c}_\ell$ is the fermionic annihilation operator at site $\ell$ and $\hat{n}_\ell=\hat{c}_\ell^\dagger \hat{c}_\ell$. Initially the left half has only one hole, while the right site has only one particle. Two detectors are placed making simultaneous stroboscopic projective measurements of particle numbers at sites $p$ and $q$ (here $p=3$, $q=5$) in intervals of time $\tau$. The `signal' is simultaneous detection on both detectors, the projection operator being $\hat{P}=\hat{n}_p\hat{n}_q$. 
\label{fig:schematic} }
\end{figure}

As a concrete example, we consider the model Hamiltonian described and schematically shown in Fig.~\ref{fig:schematic}. With $\varepsilon_0=0$, this Hamiltonian can be Jordan-Wigner transformed into the integrable XXZ qubit chain. With $\varepsilon_0>0$, the model becomes the non-integrable single impurity XXZ chain \cite{Santos_2004,Brenes_2020_1,Brenes_2020_2}, which has been of interest recently because, despite being non-integrable, it inherits the ballistic transport of the integrable XXZ chain for $\Delta<J$ at high temperatures \cite{Brenes_2018,Brenes_2020_2}. 

We divide the chain into left and right halves, the left half consisting sites $-N/2+1$ to $0$, and the right half consisting of the remaining sites. 
We consider the case where, initially, there is only one hole (i.e, there are $N/2-1$ particles) on the left half of the chain, while there is only one particle on the right half of the chain (see Fig.~\ref{fig:schematic}). Note that this does not correspond to a single configuration. The exact form of initial state will be discussed later. We put two detectors on the right half, at sites two arbitrary sites $p$ and $q$, $p,q>0$, at a finite distance from the middle. They make simultaneous projective measurements of particle number in intervals of $\tau$. We take simultaneous detection at the two sites as our signal. The corresponding projection operator is $\hat{P}=\hat{n}_p \hat{n}_{q}$, so $\hat{Q}=\hat{\mathbb{I}}-\hat{n}_p \hat{n}_{q}$. 

{\it Physics governed by $\hat{M}_Q(\tau)$ ---}
From Eq.\eqref{def_Rn_MQ}, we see that the physics of QMBDP is governed by spectral properties of  $\hat{M}_Q(\tau)$. 
The no-detection probability $R_n(\tau)$ is bounded from above by $1$. So, the spectral radius of $\hat{M}_Q(\tau)$, i.e, the highest magnitude of its eigenvalues, must be $\leq 1$. Consequently, in complete generality, we can write the eigenvalues of $\hat{M}_Q(\tau)$ as $\{e^{-\lambda_m(\tau) + i \theta_m(\tau)}\}$, with $\lambda_m\geq 0$, $\theta_m$ being real, $m$ going from $1$ to $D_Q$, where $D_Q < D$ is the Hilbert space dimension of the $\hat{Q}$ subspace. 

Let the eigenvalues of $\hat{M}_Q(\tau)$ be arranged in ascending order of $\lambda_m(\tau)$. Then, we immediately see that for $\lambda_1(\tau)>0$, i.e, when spectral radius is smaller than unity, if $n\gg 1/\lambda_1(\tau)$, the signal is almost certainly detected, irrespective of the initial state. Thus, $\tau/\lambda_1(\tau)$ gives the time scale for certainly detecting the signal. It is crucial to note that,  this finite time scale for certainly detecting the signal irrespective of the initial state arises due to repeated stroboscopic measurements,  and has no analog in absence of such measurements.

An interesting case arises if $\hat{M}_Q(\tau)$ has unit spectral radius, i.e, $\lambda_1(\tau)=0$. In this case, depending on whether the initial state has substantial overlap with the corresponding eigenvector of $\hat{M}_Q(\tau)$, there is a finite probability that the signal is never detected. For arbitrary finite $\tau$, this condition can happen if and only if some eigenvectors of $\hat{H}$ belong entirely to the $\hat{Q}$ subspace, i.e, are simultaneous eigenvectors of $\hat{Q}$ with eigenvalue $1$ \cite{supp}. 
Let the number of such eigenvectors be $D_{Q}^\prime$, $D_{Q}^\prime \leq D_Q$, and the projection operator onto this subspace be $\hat{Q}^\prime$. Then, $\hat{H}$ can be block-diagonalized as 
\begin{equation}
\label{block_diagonal_H}
\hat{H}=\hat{Q}^\prime\hat{H}\hat{Q}^\prime + \hat{P}^\prime\hat{H}\hat{P}^\prime,~~ \hat{P}^\prime=\hat{\mathbb{I}}-\hat{Q}^\prime.
\end{equation}
 If the initial state belongs to $\hat{Q}^\prime$ subspace, the Hamiltonian dynamics does not take it outside of this subspace, and hence the signal will never be detected. This is irrespective of the value of $\tau$. 
This understanding lets us relate unit spectral radius of $\hat{M}_Q(\tau)$ to specific spectral gaps in Hamiltonian.

{\it Relation with spectral gaps of Hamiltonian---} Let $\hat{H}=\hat{H}_0 + \hat{H}_1$, where $\hat{H}_0$ is a simpler Hamiltonian, whose spectral properties are easily accessible, and $\hat{H}_1$ acts as a `perturbation' on it. In particular, we consider a situation where, $D_{Q_0}$ number of eigenvectors of $\hat{H}_0$ are known to completely belong to $\hat{Q}$ subspace, with $D_{Q_0} \leq D_Q$. Let $\hat{Q}_0$ be the projection operator for this subspace. We have the block-diagonal structure $\hat{H}_0=\hat{Q}_0\hat{H}_0\hat{Q}_0 + \hat{P}_0\hat{H}_0\hat{P}_0$, $\hat{P}_0=\hat{\mathbb{I}}-\hat{Q}_0$. The Hamiltonian $\hat{H}_1$ mixes the two subspaces, but has no component completely within the $\hat{Q}_0$ subspace. The question is, with above assumptions, under what condition can a similar block-diagonalization be approximately preserved in presence of $\hat{H}_1$.

The answer is succinctly provided by VVPT. Let $|E_\alpha^{Q_0} \rangle$ be the eigenstate of $\hat{H}_0$ in $\hat{Q}_0$ subspace with energy $E_\alpha^{Q_0}$, while $|E_\nu^{P_0} \rangle$ be the eigenstate of $\hat{H}_0$ in $\hat{P}_0$ subspace with energy $E_\nu^{P_0}$. The Eq.\eqref{block_diagonal_H} is approximately satisfied if  for some range of $\alpha \in \{\alpha_{min},\alpha_{max}\}$ the eigenstates of $\hat{H}_0$ in $\hat{Q}_0$ subspace are energetically gapped from those of $\hat{P}_0$ subspace in the sense
\begin{equation}
\label{vvpt_condition}
g_\alpha\vcentcolon=\max_\nu\left|\frac{\langle E_\alpha^{Q_0} | \hat{H}_1 |E_\nu^{P_0} \rangle }{E_\alpha^{Q_0}-E_\nu^{P_0}}\right| \ll 1,~\alpha \in \{\alpha_{min},\alpha_{max}\}.
\end{equation} 
 Let the number of such eigenstates be $D_Q^\prime \leq D_{Q_0}$, and $\hat{Q}^\prime$ be the projection operator onto this subspace. Under such conditions, starting with $\hat{H}$ written in eigenbasis of $\hat{H}_0$, van Vleck perturbation theory gives a systematic way to perturbatively find a unitary operator $\hat{U}_r$ to $r$th order such that $\hat{U}_r^\dagger \hat{H} \hat{U}_r =  \hat{H}^{(r)}$ is approximately block diagonal, i.e, $\hat{H}^{(r)}\simeq \hat{Q}^\prime\hat{H}^{(r)}\hat{Q}^\prime + \hat{P}^\prime\hat{H}^{(r)}\hat{P}^\prime$, with $\hat{P}^\prime=\hat{\mathbb{I}}-\hat{Q}^\prime$ \cite{Shavitt_1980,Cohen_book}. On further diagonalizing $\hat{H}^{(r)}$ the two subspaces mix only little.  It follows that, Eq.\eqref{block_diagonal_H} is satisfied to a good approximation.

\begin{figure}[t!]
\includegraphics[width=\columnwidth]{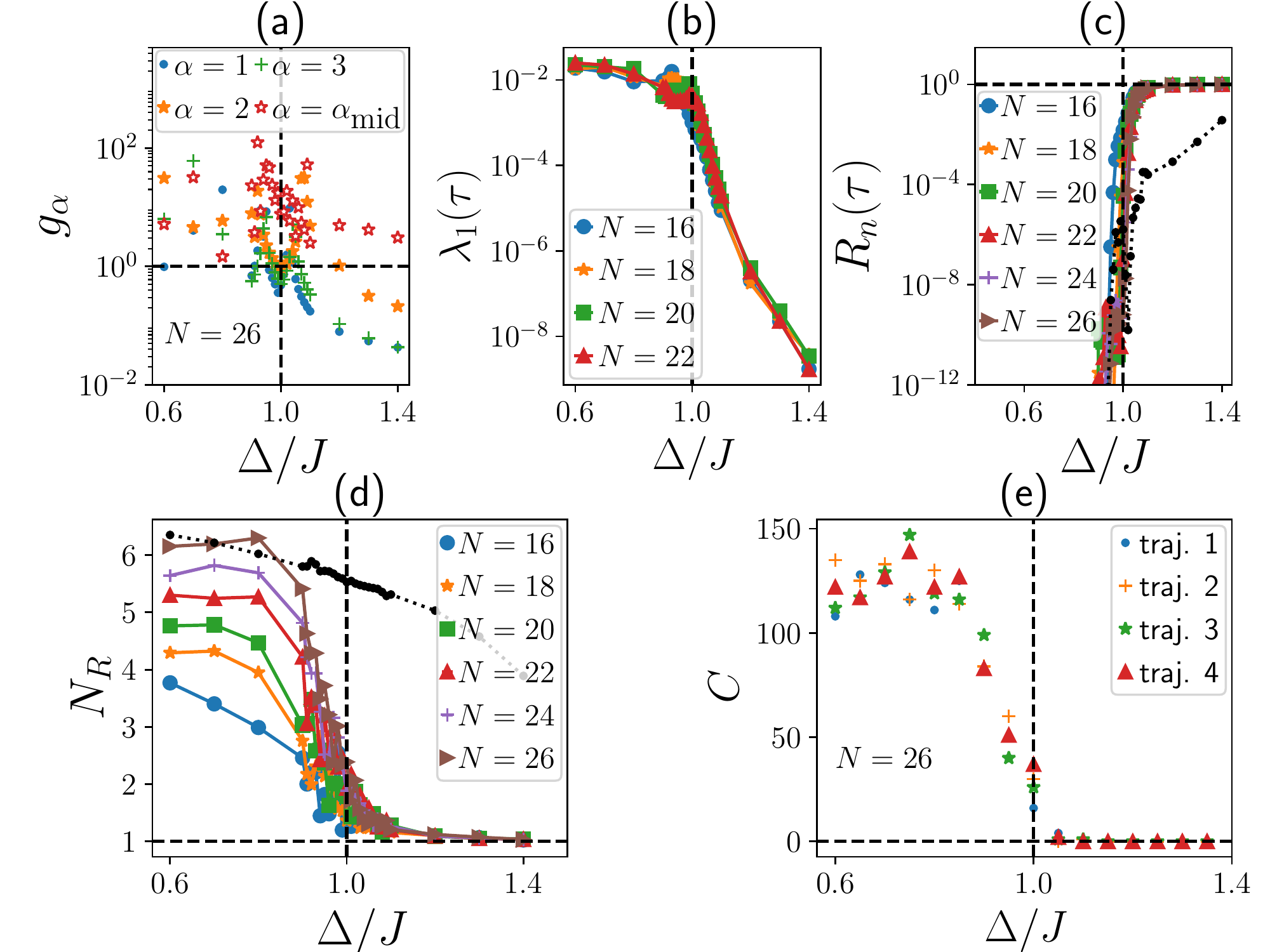}
\caption{(a) Plot of $g_\alpha$ (see Eq.\eqref{vvpt_condition}) with $\Delta$ for various values of $\alpha$. Here $\alpha_{\rm mid}=\lceil D_{Q_0}/2 \rceil$, and $N=26$. (b) Plot of $\lambda_1(\tau)$ with $\Delta$, for various system sizes. We choose $\tau=2J^{-1}$ (c) Plot of no-detection probability $R_n(\tau)$ with $\Delta$ after $n=1000$ steps. (d) Plot of $N_R$ with $\Delta$ at time $n\tau=2000J^{-1}$ in absence of any detector. For the symbols in (c) and (d) the initial state is of the form in Eq.\eqref{initial_state} with $E=E^{Q_0}_1$. The dots in (c) and (d) are the corresponding plots starting from a different initial state with $E=E^{Q_0}_{\alpha_{\rm mid}}$, and system size $N=26$. (e) The number of times $C$ the signal is detected  in $n=1000$ steps in individual runs
 of the experiment starting with initial state of the form in Eq.\eqref{initial_state} with $E=E^{Q_0}_1$ , is plotted as function of $\Delta$. The plot shows results for four different runs (trajectories). Transition at single trajectory level is clear. Other parameters, $\varepsilon_0=0.5 J$, $\sigma=0.1 J$, detector sites $p=3$, $q=5$. {\it Numerical techniques:} Panel (a) is obtained via exact diagonalization. Panel (b) is obtained via Arnoldi iteration \cite{Arnoldi_1951,Jia_2000} with sparse matrix methods  \cite{supp}, which was computationally possible up to $N=22$ (Hilbert space dimension $D=705432$). Time evolutions required for panels (b), (c), (d) and (e) are done with Chebyshev polynomial method \cite{Tal-Ezer_1984,Chen_1999,Dobrovitski_2003,Fehske_2009} with sparse matrix, which was computationally possible up to $N=26$ ($D=10400600$).
\label{fig:plots} }
\end{figure}

In our example, we choose $\hat{H}_1=-\frac{J}{2}\left(\hat{c}_{1}^\dagger \hat{c}_{0} +\hat{c}_{0}^\dagger \hat{c}_{1} \right)$, i.e, just the hopping term between left and right halves. Then $\hat{H}_0=\hat{H}-\hat{H}_1$ is the Hamiltonian without this hopping term. Without this hopping, the number of particles in the left ($N_L$) and the right ($N_R$) halves are separately conserved. Let us restrict to the half-filling case, $N_L + N_R=N/2$, so that $N_R$ is the only remaining quantum number. We immediately see that our choice of initial condition belongs to $N_R=1$ sector of $\hat{H}_0$.  We take the initial state as an energy filtered random state in this sector of $\hat{H}_0$
\begin{equation}
\label{initial_state}
\left| \psi (0) \right\rangle \propto {\rm exp}\left[-\left(\frac{\hat{H}_0-E}{\sigma}\right)^2 \right] \left | \psi_{\rm rand} \right\rangle_{N_R=1},
\end{equation}
where  the  $\left| \psi_{\rm rand} \right\rangle_{N_R=1}$ is a randomly chosen state in $N_R=1$ sector, and the prefactor is a Gaussian filter peaked around energy $E$ with a standard deviation $\sigma$.

Since our `signal' is detecting two particles simultaneously in the right half,  we define $\hat{Q}_0$ as the projector onto $N_R=0,1$ sectors. Our choice of initial state belongs to the $\hat{Q}_0$ subspace.  It is also clear that $\hat{H}_1$ connects $N_R$ and $N_R+1$ sectors. Therefore, to evaluate Eq.\eqref{vvpt_condition}, we only need to diagonalize $\hat{H}_0$ in $N_R=1$ and $N_R=2$ subspaces. The Hilbert space dimensions of $N_R=1$ and $N_R=2$ subspaces of $\hat{H}_0$ scale as $N^2$ and $N^4$ respectively, which are far smaller than than exponential scaling of the Hilbert space dimension of the full $\hat{H}$ in the half-filling sector.

In Fig.~\ref{fig:plots}(a), we plot $g_\alpha$ as a function of $\Delta$, for various values of $\alpha$. We have arranged the eigenstates in $\hat{Q}_0$ subspace such that $\alpha=0$ corresponds to $N_R=0$, which is just one configuration, and $\alpha \geq 1$ are the eigenstates in $N_R=1$ sector in ascending order of energy. In Fig.~\ref{fig:plots}(a), we see that for the lowest few eigenstates in $N_R=1$ sector, $g_\alpha \ll 1$ when $\Delta>J$, while this is not the case for $\Delta<J$.  Contrarily, for a mid spectrum state, $\alpha=\alpha_{\rm mid}=\lceil D_{Q_0}/2 \rceil$, we find $g_\alpha>1~\forall~\Delta$. Thus, we see clear evidence that Eq.\eqref{vvpt_condition} is satisfied in a low energy range when $\Delta>J$, while for $\Delta<J$, it is not satisfied. It is interesting to note that, although the Hamiltonian parameters, $J$, $\Delta$ and $\varepsilon_0$ are of same order, $g_\alpha$ emerges as a perturbative parameter for VVPT in the low-energy regime when $\Delta>J$.

{\it  Transition in detection probability---}
Given that we have a situation where on changing a parameter in $\hat{H}_0$ across some value, an energy gap opens between some eigenstates in $\hat{Q}_0$ subspace and those of $\hat{P}_0$ subspace in the sense of Eq.\eqref{vvpt_condition}, it is now clear that this will lead to a sharp decrease in $\lambda_1(\tau)$. 
Indeed, such sharp decrease in $\lambda_1(\tau)$ on tuning $\Delta/J$ across $1$
is clearly seen in Fig.~\ref{fig:plots}(b). In terms of the non-unitary matrix $\hat{M}_Q(\tau)$, this change in $\lambda_1(\tau)$ is reminiscent of gap closing in a quantum  transition. Physically, the transition is from a regime where the signal is almost certainly detected in a finite time irrespective of initial state (QMBDP $\approx 1$), to a  where, depending on the initial state, it may not be detected (QMBDP $\approx 0$), as we show next.

For large $n$, no-detection probability goes as $R_n(\tau) \sim e^{-2\lambda_1 n}$ (see Eq.\eqref{def_Rn_MQ}). So Fig.~\ref{fig:plots}(b) then suggests that, if we start from an initial state of the form of Eq.\eqref{initial_state} with $E=E^{Q_0}_1$, and fix the number of steps $n$ to be in the range $10^2 \ll n \ll 10^4$, which is finite but large, we should see a sharp transition in detection probability on tuning $\Delta/J$ across $1$. 
This is shown in Fig.~\ref{fig:plots}(c), where we plot the corresponding $R_n(\tau)$, with $n=1000$, as a function of $\Delta/J$. For all values of $N$, $R_n(\tau)$ shows an increase of more than twelve orders of magnitude on tuning $\Delta/J$ from $0.9$ to $1.1$, reaching $R_n(\tau) \approx 1$ for larger values. On increasing $N$, the transition becomes sharper, although finite-size effect is small because the detectors are placed in the bulk of the system, at a finite distance from the middle.

It is tempting to explain this transition from the known fact that  transport goes from ballistic to diffusive on going across $\Delta=J$ \cite{Bertini_2021}, which may cause particles from the left half to not reach the sites $p$ and $q$ within the chosen time. However, this would be inconsistent, because the sites $p$ and $q$ are chosen at a finite distance from the middle and the time $n\tau=2000 J^{-1}$ should have been large enough to transport particles diffusively.  Instead, our understanding in terms of VVPT consistently explains the phenomenon.

When we start with an initial state of the form of Eq.\eqref{initial_state} with $E=E^{Q_0}_{\alpha_{\rm mid}}$, we still see an exponential rise in $R_n(\tau)$ on going across $\Delta/J=1$, as shown by the dotted line in Fig.~\ref{fig:plots}(c). This is because, such a state also has a small overlap with the low energy states of $\hat{H}_0$. Nevertheless, since mid-spectrum states of $\hat{H}_0$ do not satisfy Eq.\eqref{vvpt_condition}, as seen in Fig.~\ref{fig:plots}(a), $R_n(\tau) \ll 1$ even for $\Delta/J>1$ in this case, within the range of parameters considered.

{\it Relation with domain-wall melting---}
On Jordan-Wigner transforming, our choice of initial state is akin to a domain wall in the sense that total magnetization of the left half ($\propto N_L-N/2$) is positive and that of the right half ($\propto N_R-N/2$) is negative. Our understanding in terms of VVPT says that, when starting from an initial state with $E=E^{Q^0}_1$, for $\Delta>J$, the Hamiltonian dynamics hardly takes the system out of $N_R=1$ subspace. Thus, in such a case, the initial domain wall `does not melt' up to a long time. Contrarily, for $\Delta<J$, the particles should be evenly distributed between left and right halfves, as expected in a generic system. So, the domain wall should melt.

This is clearly seen in Fig.~\ref{fig:plots}(d), where we show plots of the number of particles on the right half, $N_R$, at time $n\tau$ in absence of any detectors, i.e, for continuous time evolution with the system Hamiltonian. Thus, even in absence of any measurement, there is transition in non-equilibrium dynamics on tuning $\Delta$ across $J$. However, this transition is not as sharp as that in $R_n(\tau)$. Nevertheless,  since we find $N_R\sim 1$ for $\Delta>J$, Fig.~\ref{fig:plots}(d) establishes that putting detectors at any two sites on right half, and choosing any finite value of $\tau$, would lead to a similar transition in $R_n(\tau)$. 

Consistently with VVPT, when starting from initial state with $E=E^{Q_0}_{\alpha_{\rm mid}}$, we find that $N_R$ decreases smoothly with $\Delta$ with no hint of any transition, as shown by dotted line in Fig.~\ref{fig:plots}(d). Thus, in this case, the domain wall melts for all values of $\Delta/J$ within the observed time. This is interesting because in terms of magnetization of left and right halves, there is no difference between the two initial states with $E=E^{Q_0}_{\alpha_{\rm mid}}$ and $E=E^{Q_0}_{1}$. To our knowledge, the physics of domain wall melting has been previously explored in the single-impurity non-integrable system in only one work \cite{Santos_2011}, although it has been extensively studied for integrable XXZ chains \cite{Misguich_2019, Gamayun_2019,Misguich_2017,Stephan_2017,Mossel_2010,Yuan_2007,Gobert_2005}.

{\it Transition at single trajectory level ---}
Our understanding of the transition in terms of $\hat{M}_Q(\tau)$ shows that irrespective of the initial state, after every approximately $\sim 1/\lambda_1(\tau)$ stroboscopic measurements, the signal is detected. Let $C$ be the number of times the signal is detected in $n$ steps in a single run of the experiment. Note that $C$ is a stochastic variable, in general having different values for each run of the experiment. However, by above argument, if $n \gg 1/\lambda_1(\tau)$, which implies $R_n(\tau) \ll 1$, we expect a large value of $C$. Contrarily, if $R_n(\tau) \sim 1$, we expect a small value of $C$. Therefore, we find that a transition corresponding to that in $R_n(\tau)$ will be seen in terms of $C$ at a single trajectory level.  This is confirmed in Fig.~\ref{fig:plots}(e), where we show results for four trajectories obtained from Monte Carlo simulation \cite{supp}.

This is remarkable since observing far-for-equilibrium transitions in quantum systems usually requires measurement of expectation values as function of time. To measure expectation values at a chosen time point for a given set of system parameters, one requires averaging over measurement outcomes of several identical runs of the experiment. Each run includes preparing the initial state, evolving up to the chosen time point and making the measurement. Then, to obtain expectation values at the next time point, the entire process has to be repeated. Finally, the whole set of steps need to be repeated for several values of system parameters to obtain the transition.  Fundamentally, the requirement of having several such identical runs for each time point stems from the need to avoid effects of measurement backaction while obtaining expectation values. Instead, a transition in QMBDP takes into account the effects of measurement backaction. Consequently, as shown in Fig.~\ref{fig:plots}(e), the above transition can be seen by counting the number of simultaneous clicks in the two detectors in one single run of the experiment for each value of $\Delta/J$. This is certainly experimentally more appealing than observing the transition via dynamics of expectation values.


This also has potential technological implications. A transition in QMBDP might be useful in quantum Hamiltonian learning and parameter estimation \cite{Gebhart_2023,Yang_2022,Qin_2022,Wang_2017,Burgarth_2017,Yuan_2015,Zhang_2014,Pang_2014}. For example, in our setting, if $\Delta$ is unknown but $J$ is tunable, running the experiment only once for every value of $J$ over a wide enough range, $\Delta$ can be estimated, since the transition occurs at single trajectory level at $\Delta=J$. No ensemble averaging would be required for this. Detailed exploration of such applications of transitions in QMBDP is beyond the scope of the paper and will be carried out in future works.

{\it Conclusions---}
In conclusion, the physics of QMBDP, that we introduce and explore in this work, is experimentally relevant and both of fundamental and technological interest. It brings together several ideas from seemingly disparate fields, like statistical physics of stochastic systems, quantum measurements and quantum many-body physics, and opens the possibility of observing non-equilibrium transitions via single-shot stroboscopic measurements.
Quantum simulation experiments \cite{Schafer_2020,Morgado_2021,Monroe_2021,Daley_2022,Tran_2023} are ideal platforms to test our results. We have numerically explored one interesting example. But, the general formulation of QMBDP in Eq.~\eqref{def_Rn_MQ} and its relation to many-body spectral gaps in the sense of Eq.~\eqref{vvpt_condition}, are valid for arbitrary Hamiltonians and projection operators that define the signal. This provides the framework for future research exploring possible transitions in QMBDP in other systems and geometries, as well as for various types of signals.

{\it Acknowledgments---}
We are grateful to Klaus M\o lmer and Eli Barkai for helpful discussions. A.P acknowledges funding from the Danish National Research Foundation through the Center of Excellence ``CCQ'' (Grant agreement no.: DNRF156), Seed Grant from IIT Hyderabad, Project No.SG/IITH/F331/2023-24/SG-169. A.P acknowledges the Grendel supercomputing cluster at Aarhus University where many of the calculations were done. A.P also thanks Anupam Gupta at Indian Institute of Technology, Hyderabad, for giving access to his workstation where some of the calculations where carried out. We also thank the anonymous referees for their comprehensive reviews and constructive criticism which substantially improved the paper.

\bibliography{ref}

\newpage

\appendix

\setcounter{equation}{0}
\setcounter{figure}{0}
\renewcommand{\theequation}{S\arabic{equation}}
\renewcommand{\thefigure}{S\arabic{figure}}
\renewcommand{\thesubsection}{S\arabic{subsection}}

\section*{Supplemental Material}

\subsection{Necessary and sufficient condition for unit spectral radius of $\hat{M}_Q(\tau)$ at any finite value of $\tau$.}
\label{necessary}

{\it Sufficient---}
Let $\hat{H}$ and $\hat{Q}$ be such  that some eigenvectors of $\hat{H}$ belong entirely to the $\hat{Q}$ subspace, i.e, are simultaneous eigenvectors of $\hat{Q}$ with eigenvalue $1$. Operating $\hat{M}_Q(\tau)$, which is defined as
\begin{align}
 \hat{M}_Q(\tau) = \hat{Q} e^{-i\hat{H}\tau} \hat{Q},
\end{align}
 on such eigenvectors immediately shows that they are eigenvectors of $\hat{M}_Q(\tau)$ with eigenvalues of magnitude $1$, which proves the sufficient condition.

{\it Necessary---}
Let us define the unitary operator
$
\hat{U}_Q(\tau) = e^{-i \tau \hat{Q} \hat{H} \hat{Q} }.
$
This unitary operator is the exponential of the Hamiltonian projected to the $\hat{Q}$ subspace. We have $\hat{Q} \hat{H} \hat{Q} | E^Q_m \rangle = E^Q_m | E^Q_m \rangle$, where $E^Q_{m}$ is the $m$th eigenvalue of $\hat{Q} \hat{H} \hat{Q}$, and $| E^Q_m \rangle$ is the corresponding eigenvector. The eigenvectors span the $\hat{Q}$ subspace, and hence, $\hat{Q}| E^Q_m \rangle=| E^Q_m \rangle$. The eigenvalues of $\hat{Q}\hat{U}_Q(\tau)\hat{Q}$ are $\{e^{-i\tau E^Q_{m}}\}$, which have magnitude $1$, and corresponding eigenvectors are $\{| E^Q_m \rangle\}$. The operator $\hat{M}_Q(\tau)$ can now be written as
\begin{align}
\label{MQUQ}
\hat{M}_Q(\tau) = \hat{Q} \Big[\hat{U}_Q(\tau) + \sum_{p=2}^{\infty} \frac{(-i)^p \tau^p}{p!} \left\{ \hat{H}^p - \left(\hat{Q}\hat{H}\hat{Q}\right)^p\right\} \Big] \hat{Q}
\end{align}
Let us assume $\tau$ is small, such that $\hat{M}_Q(\tau)$ can be treated as a perturbation over $\hat{Q}\hat{U}_Q(\tau)\hat{Q}$. Keeping only the leading order of the second term on right-hand-side in above equation, we have
\begin{align}
\hat{M}_Q (\tau) \simeq \hat{Q}\hat{U}_Q(\tau)\hat{Q}-\frac{\tau^2}{2}\left( \hat{Q} \hat{H}^2 \hat{Q} -  \hat{Q} \hat{H} \hat{Q} \hat{H} \hat{Q}\right).
\end{align}
Assuming order by order expansion of eigenvalues and eigenvectors of $\hat{M}_Q(\tau)$ in $\tau$, the leading order difference between eigenvalues of $\hat{M}_Q (\tau)$ and $\hat{Q}\hat{U}_Q(\tau)\hat{Q}$ is given by expectation value of the second term in above equation, in the eigenbasis of 
$\hat{Q} \hat{H} \hat{Q}$. This yields
\begin{align}
e^{-\lambda_m(\tau)+ i\theta_m(\tau)} \simeq e^{-i\tau E^Q_{m}} - \frac{\tau^2}{2}\left[ \langle E^Q_m | \hat{H}^2 | E^Q_m \rangle - \left(E^{Q}_{m}\right)^2 \right].
\end{align}
Taking logarithm of absolute value of above equation and keeping terms only up to the leading order in $\tau$, we find the following insightful expression for $\lambda_m(\tau)$ to leading order in $\tau$, 
\begin{align}
\lambda_m(\tau)=\tau^2\left[ \langle E^Q_m | \hat{H}^2 | E^Q_m \rangle - \left(E^{Q}_{m}\right)^2 \right] + O(\tau^3).
\end{align}
This expression immediately shows that with $\tau\to 0$, $\lambda_m(\tau)\to 0$, $\forall m$. Thus, magnitude of all eigenvalues of $\hat{M}_Q(\tau)$ tend to $1$ in this limit, meaning that the signal is never detected if the detectors are always kept on. This is the standard quantum Zeno effect. 

Next, since $\lambda_m(\tau)\geq 0$, we have $\langle E^Q_m | \hat{H}^2 | E^Q_m \rangle - \left(E^{Q}_{m}\right)^2 \geq 0$. Consequently, for finite  but small $\tau$, the signal is always eventually detected, unless, for some choice of $m$, we have $\langle E^Q_m | \hat{H}^2 | E^Q_m \rangle - \left(E^{Q}_{m}\right)^2 = 0$. The later condition can only happen if  $| E^Q_m \rangle$ is an eigenvector of $\hat{H}$ with eigenvalue $E^Q_m$. Therefore, this eigenvector of $\hat{H}$ must belong completely to the $\hat{Q}$ subspace.  However, if this is the case, by multiplying Eq.\eqref{MQUQ} with  $| E^Q_m \rangle$ from right, we directly see that it is an eigenvector of $\hat{M}_Q$, the corresponding eigenvalue having $\lambda_m(\tau)=0$ irrespective of the value of $\tau$.
 Thus, some eigenvectors of $\hat{H}$ belonging completely to the $\hat{Q}$ subspace is a necessary condition for unit spectral radius of $\hat{M}_Q(\tau)$ for any finite value of $\tau$.

\subsection{System and initial state}
For numerical exploration, our system Hamiltonian is
\begin{align}
\hat{H}=-\hspace*{-15pt}\sum_{\ell=-N/2+1}^{N/2-1} \left[\frac{J}{2}\left( \hat{c}_\ell^\dagger \hat{c}_{\ell + 1} +  \hat{c}_{\ell + 1}^\dagger \hat{c}_\ell \right) + \Delta \hat{n}_\ell \hat{n}_{\ell + 1} \right]+\varepsilon_0 \hat{n}_{0}, 
\end{align}
where $\hat{c}_\ell$ is the fermionic annihilation operator at site $\ell$ and $\hat{n}_\ell=\hat{c}_\ell^\dagger \hat{c}_\ell$. We choose
\begin{align}
\hat{H}_1=-\frac{J}{2}\left(\hat{c}_{1}^\dagger \hat{c}_{0} +\hat{c}_{0}^\dagger \hat{c}_{1} \right),~~\hat{H}_0=\hat{H}-\hat{H}_1.
\end{align}
We take the initial state as an energy filtered random state in this sector of $\hat{H}_0$
\begin{equation}
\label{initial_state_sup}
\left| \psi (0) \right\rangle \propto {\rm exp}\left[-\left(\frac{\hat{H}_0-E}{\sigma}\right)^2 \right] \left | \psi_{\rm rand} \right\rangle_{N_R=1},
\end{equation}
where  the  $\left| \psi_{\rm rand} \right\rangle_{N_R=1}$ is a randomly chosen state in $N_R=1$ sector, and the prefactor is a Gaussian filter peaked around energy $E$ with a standard deviation $\sigma$.

\subsection{Numerical techniques}

{\it For dynamics and no-detection probability----}
For time-evolution of a state vector with Hamiltonian, we use Chebyshev expansion \cite{Tal-Ezer_1984,Chen_1999,Dobrovitski_2003,Fehske_2009} and carry out required matrix-vector multiplications using sparse representation of $\hat{H}$. This allowed exploring dynamics up to $N=26$, i.e, Hilbert space dimension $D=10400600$ at half-filling. For pure initial state belonging to $\hat{Q}$ subpsace, $R_n=|\langle \psi_{R_n} | \psi_{R_n} \rangle|^2$, $\psi_{R_n}=\hat{M}^n_Q(\tau) | \psi(0) \rangle$. Since $\hat{Q}=\hat{\mathbb{I}}-\hat{n}_p \hat{n}_q$ is a strictly local operator, its operation on the state vector is simple.  This allows us to operate $\hat{M}_Q(\tau)$ on the state. Repeated operations of $\hat{M}_Q(\tau)$ yield $R_n$.

{\it For calculating $\lambda_1(\tau)$---}
  We combine the above procedure for operating $\hat{M}_Q(\tau)$ on the state with the Arnoldi procedure \cite{Arnoldi_1951,Jia_2000}, to estimate the largest magnitude eigenvalue of $\hat{M}_Q(\tau)$, and thereby $\lambda_1(\tau)$. However, this becomes more difficult when $\lambda_1(\tau)$ is small and due to exponentially growing Hilbert space dimension, our results for $\lambda_1(\tau)$  are limited up to system size $N=22$ (Hilbert space dimension $D=705432$ at half-filling). 

{\it For single-trajectory simulation---}
For this purpose, we directly simulate the action of the two detectors on the dynamics as follows. The detectors makes stroboscopic measurements in steps of time $\tau$. Just before the $n$th stroboscopic measurement, let the state of the system be $|\psi_b(n\tau)\rangle$.
For detectors placed at sites $p$ and $q$, we calculate the expectation value 
\begin{align}
& \langle\hat{n}_p(n\tau)\hat{n}_q(n\tau)\rangle=\langle \psi_b(n\tau)| \hat{n}_p\hat{n}_q | \psi_b(n\tau) \rangle 
\end{align}
This gives the probability of simultaneous detection at $p$ and $q$ sites. To simulate the action of the detectors, we choose an independent random numbers, $r$ from a uniform distribution between $0$ and $1$, and obtain $| \psi_f(n\tau) \rangle$ as follows,
\begin{align}
& {\rm if}~~r\leq \langle\hat{n}_p(n\tau)\hat{n}_q(n\tau)\rangle: \nonumber \\
&\qquad  | \psi_f(n\tau) \rangle = \hat{n}_p\hat{n}_q | \psi_b(n\tau) \rangle,~~C\to C+1 \nonumber \\
& {\rm else}:  \nonumber \\
&\qquad | \psi_f(n\tau) \rangle=| \psi_b(n\tau) \rangle-\hat{n}_p\hat{n}_q | \psi_b(n\tau) \rangle
\end{align}
In above $C$ is a counter variable that is increased by one if there is simultaneous detection, i.e, if the first condition above is satisfied. After the above operation the state $| \psi_f(n\tau) \rangle$ is normalized and evolved with the system Hamiltonian for a time $\tau$ before the next stroboscopic measurement is simulated similarly. It is clear from above that $C$ is a stochastic variable, every run of the simulation would in general yield a different value of $C$. The value of $C$ gives how many times the `signal' is detected in a single run of the experiment.

\subsection{Dynamics}
\label{dynamics}

\begin{figure}
\includegraphics[width=\columnwidth]{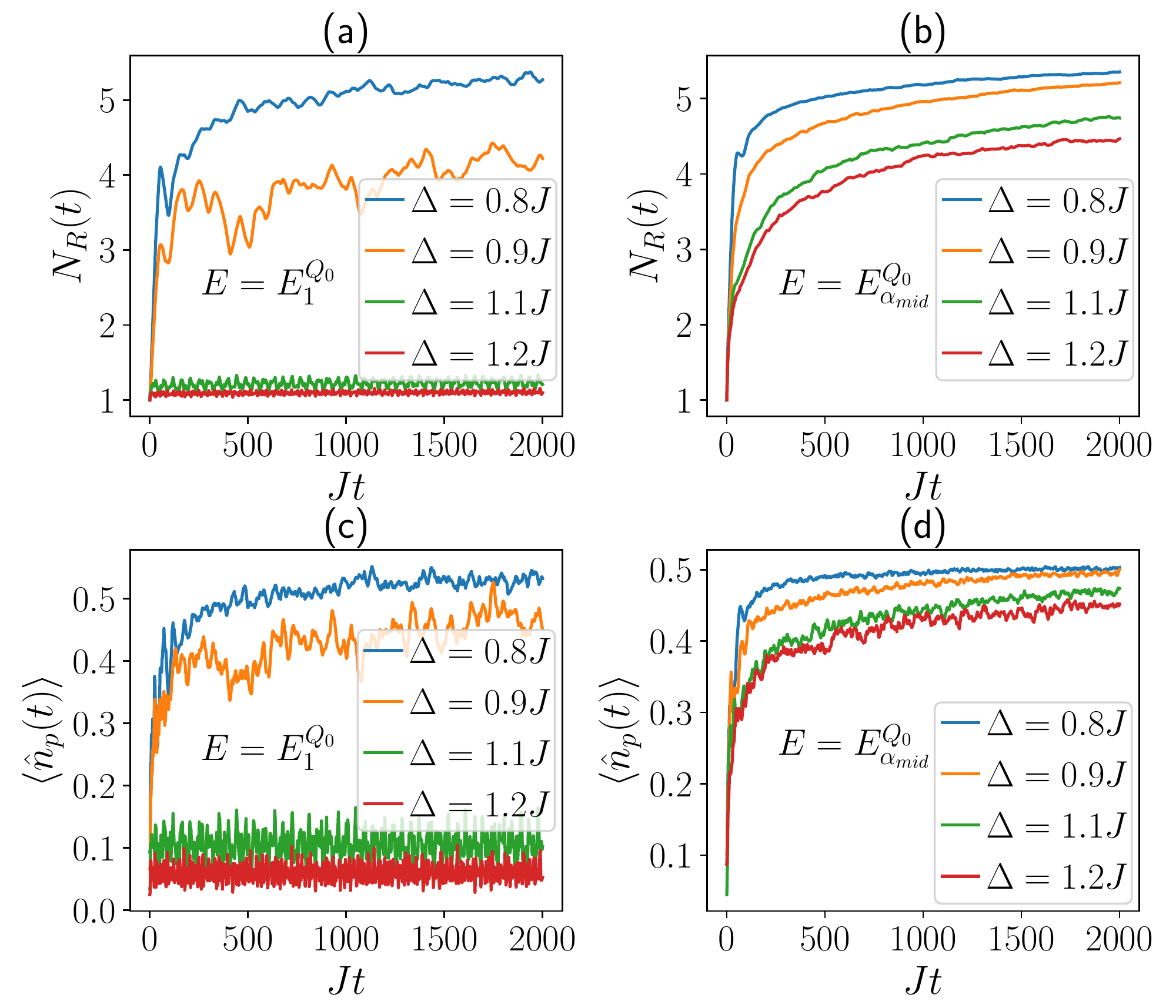}
\caption{(a) Dynamics of number of particles in right half, $N_R(t)$, for various values of $\Delta$, when starting from an initial state of the form Eq.\eqref{initial_state_sup} with $E=E_1^{Q_0}$. (b) The same with $E=E^{Q_0}_{\alpha_{\rm mid}}$. (c) Dynamics of occupation at site $p=3$, for various values of $\Delta$, when  when initial state is chosen with $E=E^{Q_0}_1$ in Eq.(4) of main text. (d) The same with $E=E^{Q_0}_{\alpha_{\rm mid}}$. Other parameters, $N=22$, $\epsilon_0=0.5J$, $\sigma=0.1 J$.
\label{fig:dynamics} }
\end{figure}

Here we show the plots of dynamics of the system, in absence of any measurements, starting from the two different initial states corresponding to $E=E^{Q_0}_{1}$, $E=E^{Q_0}_{\alpha_{\rm mid}}$. The plots are shown in Fig.~\ref{fig:dynamics}. 

In Fig.~\ref{fig:dynamics}(a), when $E=E^{Q_0}_{1}$, we see that for $\Delta<J$, $N_R(t)$ (where we now explicitly write the time argument) increases with $t$ towards $N_R(t) \sim N/2$. For $\Delta>J$, $N_R(t)$ remains close to $N_R(t)\sim 1$, over the entire time range considered. When $E=E^{Q_0}_{\alpha_{\rm mid}}$, $N_R(t)$ increases towards $N/2$ for all values of $\Delta$, an shown in Fig.~\ref{fig:dynamics}(b). So, in the domain wall picture,  we clearly see that the domain wall does not melt for $\Delta>J$ and $E=E^{Q_0}_{1}$, while in other cases, it melts, as expected in non-integrable systems.

In Fig.~\ref{fig:dynamics}(c), we plot the dynamics of occupation at site $p$, $\langle\hat{n}_p(t)\rangle$, with $p=3$, for various values of $\Delta$, starting with the initial state corresponding to  $E=E^{Q_0}_{1}$. The occupation approaches $\langle\hat{n}_p(t)\rangle \sim 0.5$ for $\Delta<J$, while it shows oscillations about a much smaller value for $\Delta>J$. The important point to note here is that even for $\Delta>J$, where the domain wall does not melt, the dynamics is not completely frozen. Rather, it is restricted to a small subspace. In Fig.~\ref{fig:dynamics}(d), we show similar plots starting with the initial state corresponding to $E=E^{Q_0}_{\alpha_{\rm mid}}$. In this case, the occupation approaches $\langle\hat{n}_p(t)\rangle \sim 0.5$ for all values of $\Delta$ considered here.

\subsection{Single detection after time $n\tau$}
\begin{figure}
\includegraphics[width=\columnwidth]{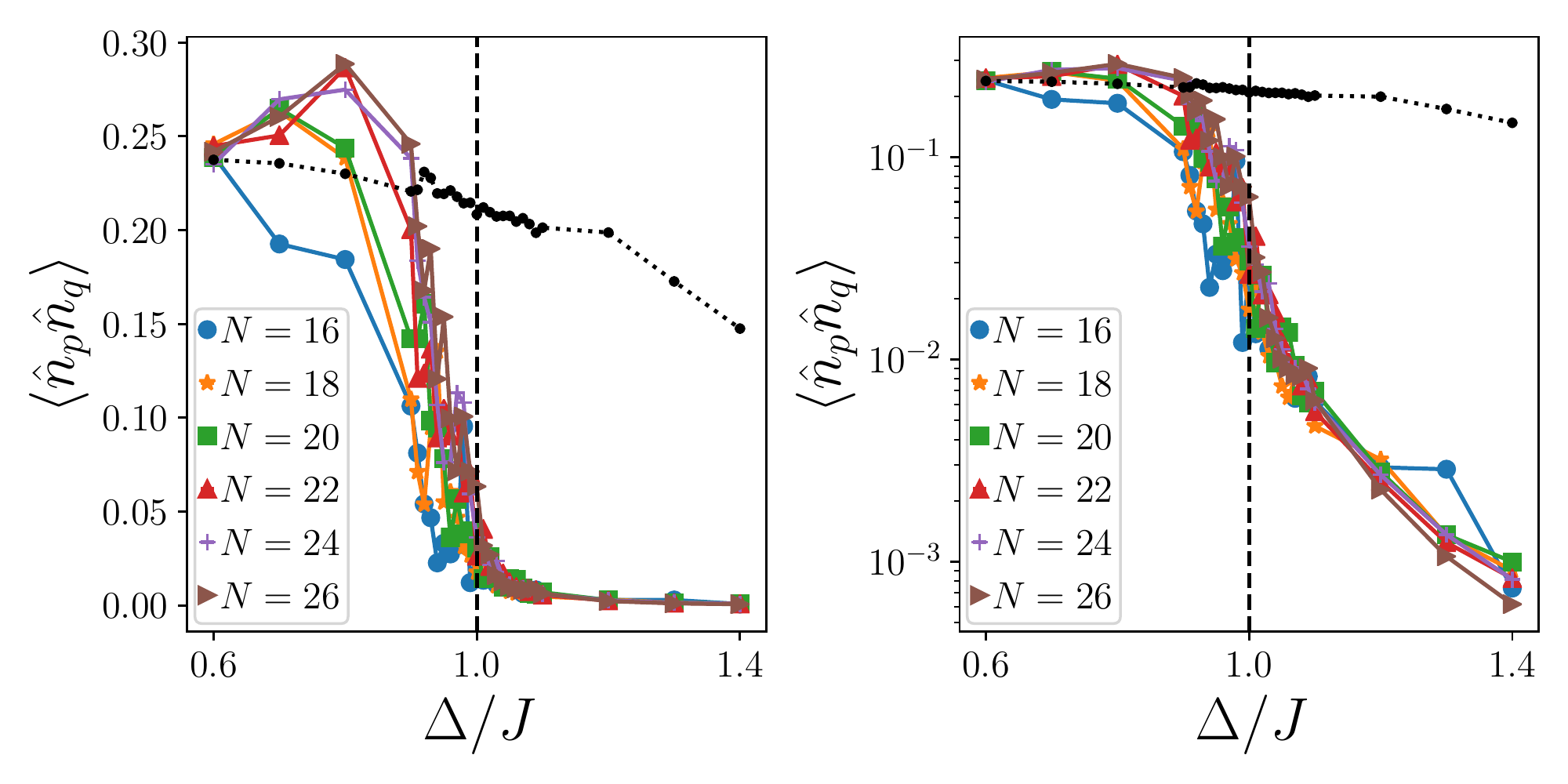}
\caption{Panel (a) shows plot of the expectation value $\langle \hat{n}_p \hat{n}_q \rangle$ at time $2000J^{-1}$ versus $\Delta/J$ for various system sizes, when starting from an initial state of the form Eq.\eqref{initial_state_sup} with $E=E_1^{Q_0}$. The black dots shows the same for $N=26$ when starting from an initial state of the form Eq.\eqref{initial_state_sup} with $E=E_{\alpha_{\rm mid}}^{Q_0}$. Panel (b) shows the same plot with y-axis in log scale.  This expectation value gives the probability of getting simultaneous clicks on two detectors at sites $p$ and $q$ in a single measurement done at time $n\tau$. Parameters: $p=3$, $q=5$, $\varepsilon_0=0.5 J$, $\sigma=0.1 J$. \label{fig:npnq} }
\end{figure}

Instead of stroboscopic measurements at steps of time $\tau$, we can also ask what is the probability of simultaneous detection at sites $p$ and $q$ in a single measurement after a given time. This is given by the expectation value $\langle \hat{n}_p \hat{n}_q \rangle$ calculated at that time. In Fig.~\ref{fig:npnq}, we show plots of this expectation value at time $2000J^{-1}$ as a function of $\Delta/J$. When starting with an initial state of the form Eq.\eqref{initial_state_sup} with $E=E_1^{Q_0}$, we see that $\langle \hat{n}_p \hat{n}_q \rangle$ shows a smooth crossover from a finite value to a small value on going across $\Delta/J=1$. Clearly, the probability of simultaneous detection on both sites, i.e, of detecting the `signal', is not close to $1$ even when  $\Delta/J<1$, within the chosen parameter regime. There is still considerable probability of not detecting the signal ($>70\%$ in the chosen parameter regime). This is very different from stroboscopic measurements where the no-detection probability was $<10^{-12}$ for $\Delta/J<0.9$, and sharply rises to $\approx 1$ for $\Delta/J=1.1$. Thus, it is clear that backaction of stroboscopic measurements makes the transition much sharper.

\begin{figure}
\includegraphics[width=\columnwidth]{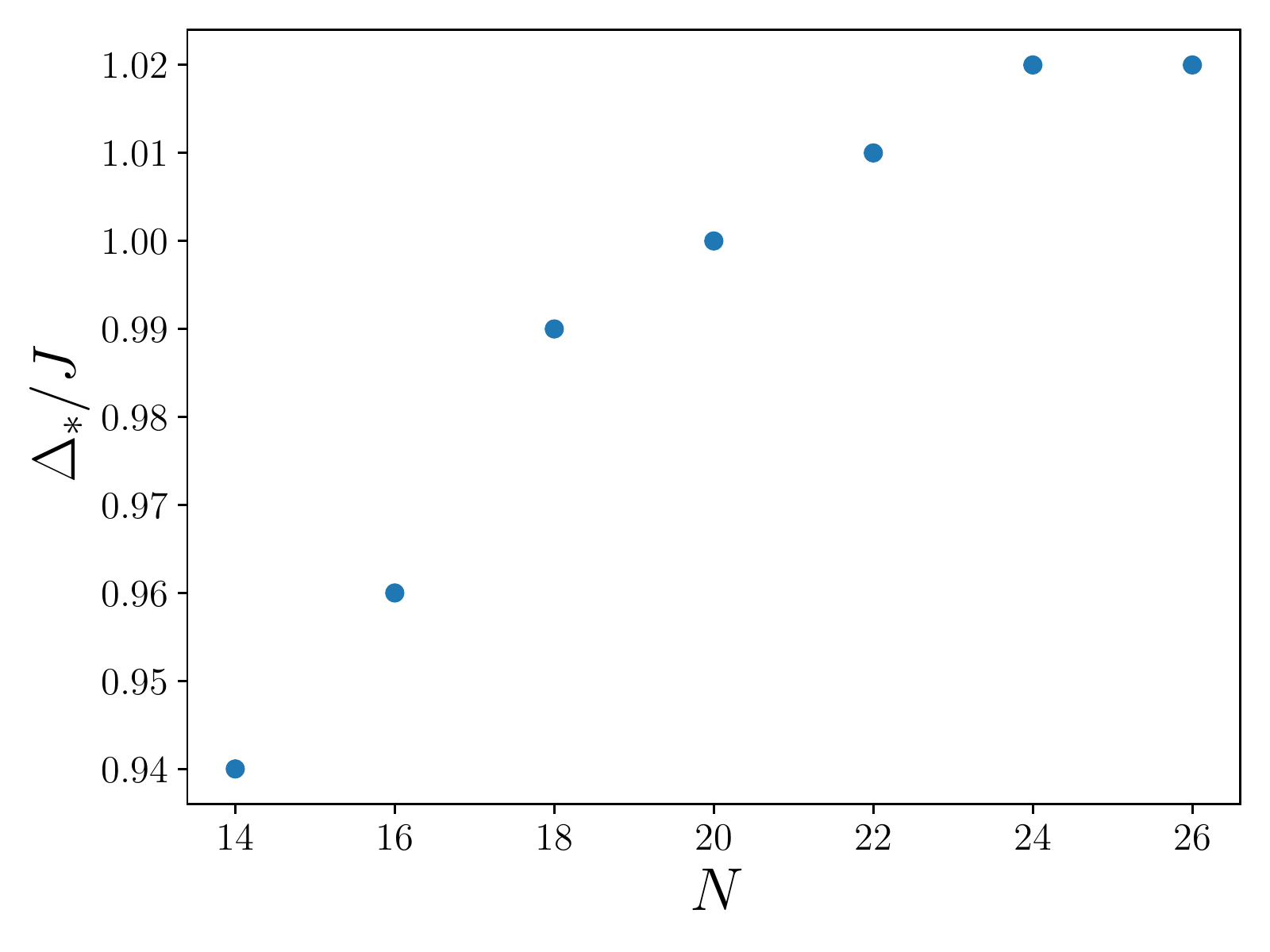}
\caption{The figure shows that value of $\Delta_*/J$ as a function of system size $N$. Parameters, $E=E^{Q_0}_1$, $\epsilon=10^{-5}$, $n=1000$, $\varepsilon_0=0.5 J$, $\sigma=0.1 J$. \label{fig:Delta_trans_vs_N} }
\end{figure}

\subsection{Behavior versus $N$}
We now show the effect of changing system sizes on the sharp transition in $R_n$. For the initial state of the form Eq.\eqref{initial_state_sup} with $E=E_1^{Q_0}$, we define $\Delta_*/J$ has the smallest value of $\Delta/J$ for which $R_n>\epsilon$, where $\epsilon$ is some chosen tolerance. In Fig.~\ref{fig:Delta_trans_vs_N}, we show the plot for $n=1000$ and $\epsilon=10^{-5}$. We see that the value of $\Delta_*/J$ changes only a little on nearly doubling the system size, and seems to saturate for larger system size. This gives further numerical evidence for the sharpness of the transition  in $R_n$.

\begin{figure}
\includegraphics[width=\columnwidth]{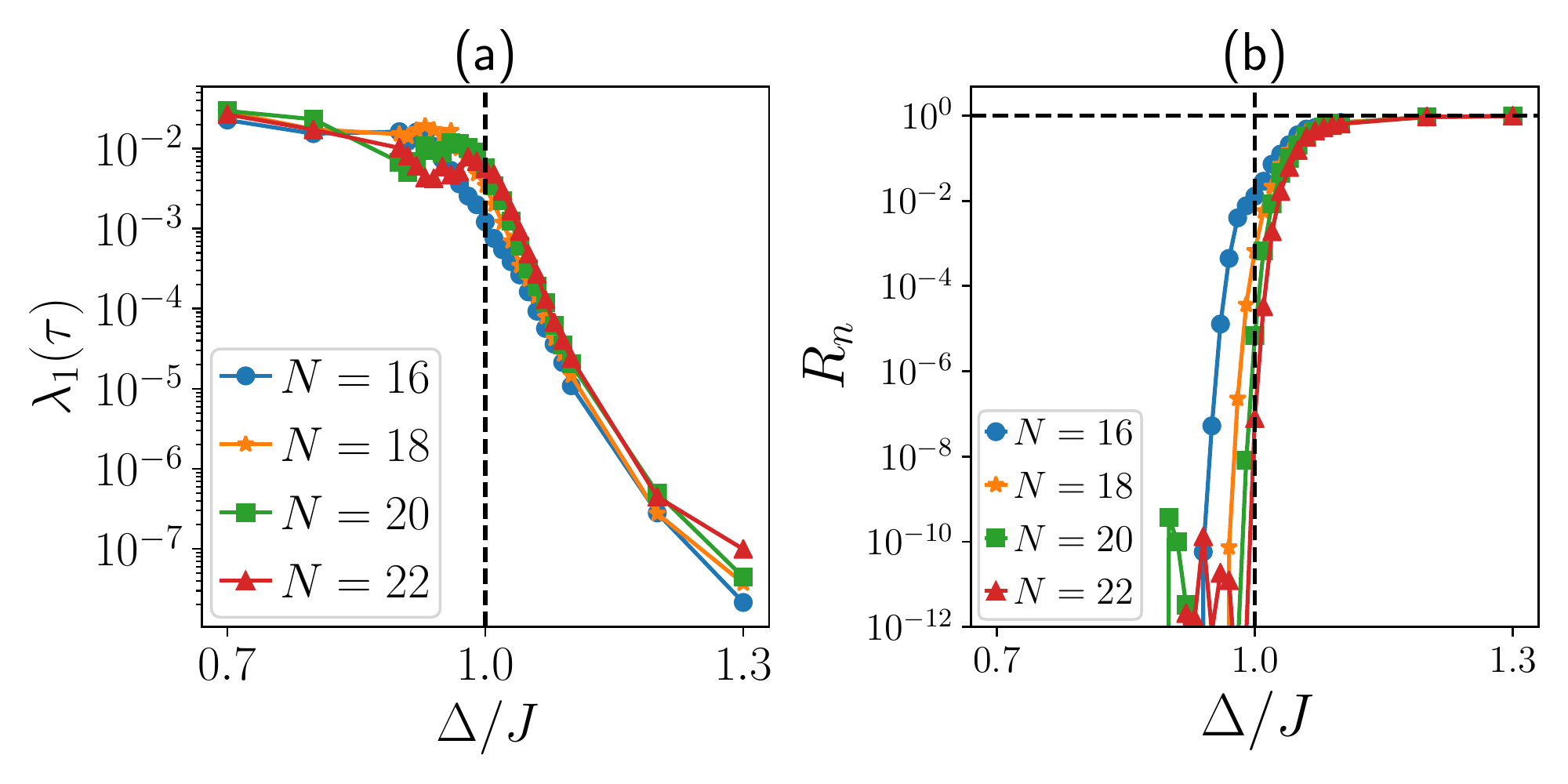}
\caption{a) Plot of $\lambda_1(\tau)$ with $\Delta$, for various system sizes, and for $\tau=4J^{-1}$. (b) Plot of no-detection probability, $R_n$, with $\Delta$ after $n=1000$ steps, for the same system sizes, and for  $\tau=4J^{-1}$. Other parameters: $E=E^{Q_0}_1$, $\varepsilon_0=0.5 J$, $\sigma=0.1 J$. \label{fig:different_tau}}
\end{figure}

\subsection{Different choice of $\tau$}
Our analytical understanding, as well as our numerical results on domain-wall melting, shows that the sharp transition in QMBDP is independent of the choice of $\tau$. In Fig.~\ref{fig:different_tau} we show plots of  $\lambda_1(\tau)$ versus $\Delta$, and no-detection probability $R_n$ for $n=1000$, and  $\tau=4J^{-1}$. This value of $\tau$ is different from that of the plots in the main text, where $\tau=2J^{-1}$. We still see the same behavior, no-detection probability changes from $R_n \sim 10^{-12}$  to $R_n \sim 1$ on changing $\Delta/J$ from $0.9$ to $1.1$.

\begin{figure}[t!]
\centering
\includegraphics[width=\columnwidth]{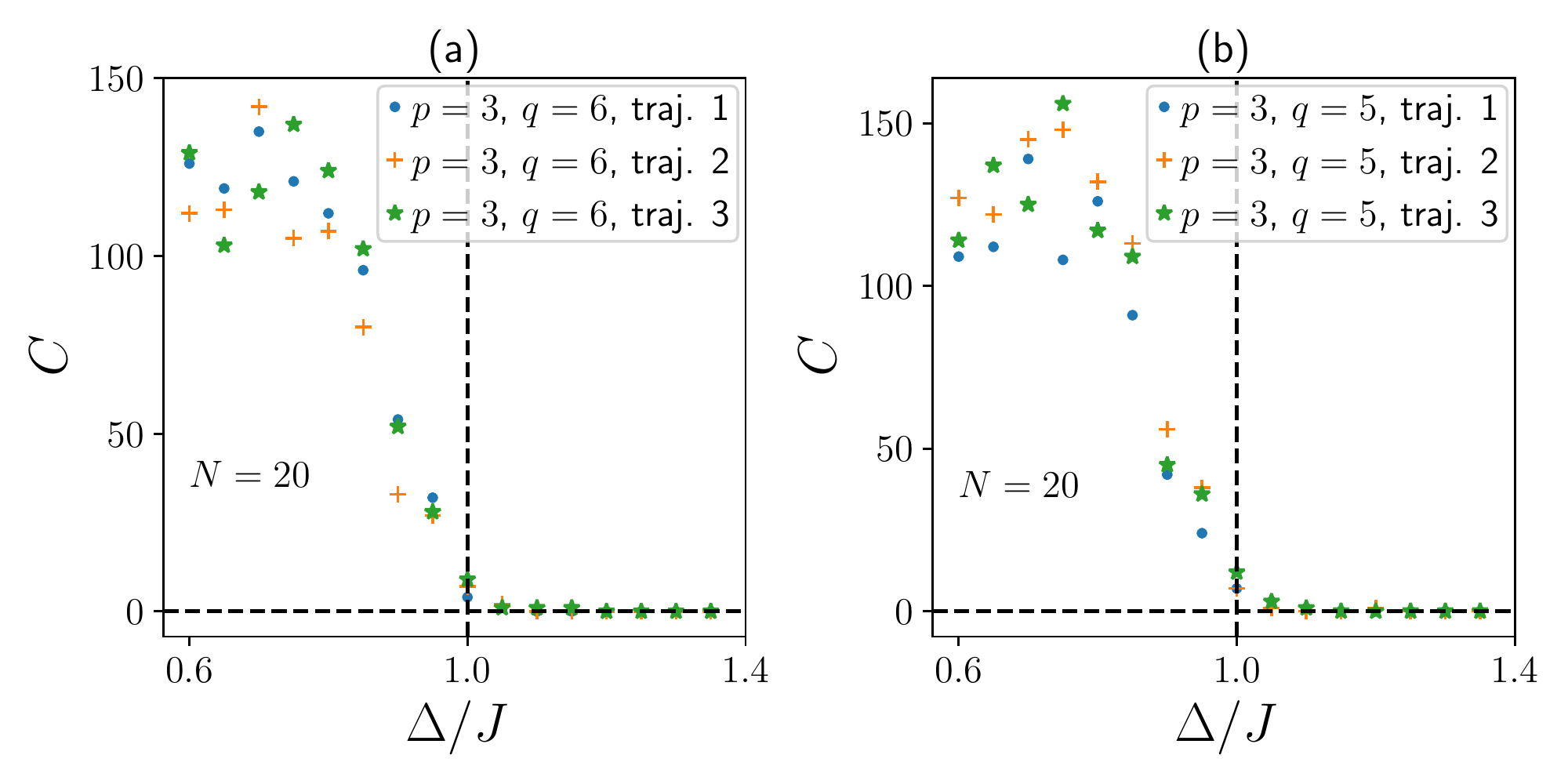}
\caption{Comparison of single trajectory plots showing the transition in cases where the two detectors are placed at (a) $p=3$, $q=6$, (b) $p=3$, $q=5$. Here $C$ is the number of simultaneous clicks in the two detectors after $n=1024$ single-shot stroboscopic measurements. Three trajectories, i.e, simulation of three runs of the experiment, are shown for each case. Other parameters, $N=20$, $\epsilon_0=0.5 J$, $\tau=2J^{-1}$.  \label{fig:different_detector_positions}}
\end{figure}

\subsection{Different position of detectors}
All numerical results for the transition in QMBDP presented till now are for the case where the detectors are placed at sites $p=3$ and $q=5$. Our analytical understanding, as well as our numerical results on domain-wall melting show that they remain unchanged for any positions of the detectors, at a finite distance from the middle. To explicitly show this, in Fig.~\ref{fig:different_detector_positions}, we plot the number of simultaneous clicks in single trajectory simulations of up to $n=1024$ stroboscopic measurements for the cases with detectors at $p=3,q=6$, and $p=3,q=5$, as function of $\Delta/J$. We clearly see identical behavior, with transition at $\Delta/J=1$.

\end{document}